\documentclass{article}
\usepackage{amsthm}
\usepackage{graphicx}
\usepackage[bottom]{footmisc}

\begin{document}

\title{A Simple Abstraction for Data Modeling}

\author{
Nassib Nassar\\
       RENCI, University of North Carolina at Chapel Hill\\
       nassar@renci.org
}

\date{October 4, 2010}

\maketitle
\begin{abstract}
The problems that scientists face in creating well designed databases intersect with the concerns of data curation.  Entity-relationship modeling and its variants have been the basis of most relational data modeling for decades.  However, these abstractions and the relational model itself are intricate and have proved not to be very accessible among scientists with limited resources for data management.  This paper explores one aspect of relational data models, the meaning of foreign key relationships.  We observe that a foreign key produces a table relationship that generally references either an entity or repeating attributes.  This paper proposes constructing foreign keys based on these two cases, and suggests that the method promotes intuitive data modeling and normalization.
\end{abstract}

\section{Introduction}
Science has become dependent on the ability to share and reuse data sets that can be very large in scale and quite heterogeneous\cite{bell09}.  In scientific communities and in digital library communities, which are concerned with dissemination and preservation of scientific and scholarly output, there has been a growing interest in many aspects of data management.  In this context the term, ``data curation,'' includes the preparation of data in order to make them reusable by others.

Data curation begins ideally when data are first produced rather than as an afterthought, and it implies an improvement in data integrity and, therefore, in the quality of data models.  One of the major contributions of database research has been the theory of normalization\cite{codd71}, which is essential to the creation of robust data models.  Normalization is a data modeling activity often performed by database specialists, but many database users find that the theory is not easy to apply.

Since the 1970s there has also been a lot of work on layering conceptual and semantic abstractions over relational databases, with extensions or mappings to the relational model or with data modeling techniques.  Entity-relationship modeling\cite{chen76} is the most frequently used conceptual modeling technique\cite{davies06} and has given rise to many variations.  These approaches allow users to systematize their knowledge about data at a conceptual level but often involve significant complexity\cite{tavana07}.  There is also a natural tension between abstract conceptual models, which obscure the underlying database, and lower-level ``logical'' models, which require a fairly sophisticated knowledge of relational modeling.

Many scientists avoid these difficulties by using spreadsheets or text files instead of databases.  The files are often extended ad hoc to encode additional structure within a table.  This leads to a variety of table-based and hierarchical data models, many of which proliferate unnormalized data.  It seems unlikely that small research labs will adopt the complicated data modeling processes that have evolved in the database and business communities.  It is possible to offer them services that can help with database design; however, because scientific research tends to evolve and branch out continually, data modeling in these domains often continues as an iterative process throughout the life of a research program.  There appears to be a need for simpler methods and tools that would allow scientists to integrate data modeling into their research process.

While the importance of data curation is becoming recognized, there is a recent movement away from relational databases and normalized data models in order to cope with massive data scaling requirements of many current applications.  Relational databases have always been far from ubiquitous in the sciences; nevertheless this most recent wave of heterogeneous database models is a major shift in the landscape.  It suggests that databases will need to interoperate with each other in various hybrid relational/non-relational information systems, at the same time as they have to interface with a much broader range of users.

In this paper a simple data modeling abstraction is proposed in the hope of making database design, and by extension data curation, more accessible to the increasing number of users that have to manage data.

\section{Definitions}
A \emph{table design} is the set of attributes $\{a_1, \ldots ,a_n\}$ represented in a database table.  In this paper we will often use ``table'' as a shorthand to mean table design.  It will be assumed that every table has a primary key, which is underlined in these examples.  We start with a traditional employee example:
\begin{eqnarray*}
E & = & \{ \underline{emp}, ediv, edept \} \\
D & = & \{ \underline{div, dept}, addr \}
\end{eqnarray*}
Table~$E$ represents the employees that work in a company, with each tuple having the name of an employee ($emp$) and the name of the division ($ediv$) and department ($edept$) that the employee works in.  Table $D$ represents the departments in the company, with each tuple having the name of a department ($dept$) and the division it is contained in ($div$), as well as the department's address ($addr$).  Since a modern company might be the result of several mergers, a department name may not be unique within the company; therefore the division and department names are used together to identify a department.

This example will also have a foreign key constraint $empdept$ from $\{ediv,$ $edept\}$ in $E$ to $\{div, dept\}$ in $D$, which will be notated as follows.  Let $S$ and $T$ be tables, and $A \subseteq S$.  Let $K_T$ be a candidate key of $T$.  A foreign key constraint $F$ from $A$ (in $S$) to $K_T$ (in $T$) is written as $F = \phi_{S,T}(A, K_T)$.

In our case we have:
\begin{eqnarray*}
empdept & = & \phi_{E,D}(\{ediv, edept\}, \{div, dept\})
\end{eqnarray*}

Most of the foreign key constraints described in this paper reference primary keys, and it will be convenient to give them a separate definition.
\theoremstyle{definition}
\newtheorem{definition}{Definition}
\begin{definition}
Let $S$ and $T$ be tables, with $A \subseteq S$.  Let $\Pi_T$ be the primary key of $T$.  If there exists a foreign key constraint $F = \phi_{S,T}(A, \Pi_T)$, then $F = \Phi_{S,T}(A)$ is a \emph{simple foreign key constraint}.
\end{definition}

Tables $E$ and $D$ are examples of a table which contains no foreign keys as part of its primary key.
\begin{definition}
Let $S$ be a table with primary key $\Pi_S$.  $S$ is an \emph{entity table} if there exists no foreign key constraint $\phi_{S,T}(A, K)$ for any $T$, $K$, or $A \subseteq \Pi_S$.
\end{definition}

Suppose that each employee can have several phone numbers and that for this purpose there is defined a new table $P$ and a foreign key constraint $empphone$:
\begin{eqnarray*}
P & = & \{ \underline{emp, phone} \} \\
empphone & = & \Phi_{P,E}(\{emp\})
\end{eqnarray*}
$P$ is another special class of table which has a foreign key as a proper subset of its primary key such that the foreign key references the primary key of another table.
\begin{definition}
Let $T$ be a table with primary key $\Pi_T$.  Then $T$ is a \emph{multivalued table} if there exists a simple foreign key constraint $\Phi_{T,S}(A)$ for some $S$ and $A \subset \Pi_T$.
\end{definition}
\begin{definition}
A table is a \emph{simple table} if it is an entity table or a multivalued table.
\end{definition}
Two operators are defined for describing referential constraints involving simple tables.
\begin{definition}
Let $S$ be a simple table, and let $T$ be an entity table.  If there exists a simple foreign key constraint $F = \Phi_{S,T}(A)$ for some $A \subset S$, then there is an \emph{entity reference} \emph{from} $S$ \emph{to} $T$, $\varepsilon(F, T)$.
\end{definition}
\begin{definition}
Let $T$ be a multivalued table with primary key $\Pi_T$.  Therefore there exists a simple foreign key constraint $F = \Phi_{T,S}(A)$ for some $S$ and $A \subset \Pi_T$.  If $S$ is a simple table, then there is a \emph{multivalued reference} \emph{from} $S$ \emph{to} $T$, $\mu(F, T)$.
\end{definition}

In our example we will include the set of entity references and multivalued references from each table.  To summarize:
\begin{eqnarray*}
E & = & \{ \underline{emp}, ediv, edept \} \\
E_R & = & \{ \varepsilon(empdept, D), \mu(empphone, P) \} \\
D & = & \{ \underline{div, dept}, addr \} \\
D_R & = & \{ \} \\
P & = & \{ \underline{emp, phone} \} \\
P_R & = & \{ \varepsilon(empphone, E) \} \\
empdept & = & \Phi_{E,D}(\{ediv, edept\}) \\
empphone & = & \Phi_{P,E}(\{emp\})
\end{eqnarray*}

Note that a multivalued reference may have a corresponding entity reference arising out of the same constraint; for example, $\varepsilon(empphone, E)$.
\begin{definition}
Let $S$ be an entity table, let $T$ be a multivalued table, and let $r_\mu = \mu(F, T)$ be a multivalued reference from $S$.  Since this implies that there exists an entity reference from $T$, $r_\varepsilon = \varepsilon(F, S)$, we call $r_\varepsilon$ a \emph{redundant entity reference} corresponding to $r_\mu$.
\end{definition}

\begin{definition}
If $T$ is a table and $R$ is the set of entity references and multivalued references from $T$, then we define an \emph{abstract table} $A(T) = (T', R')$ where $T' \subseteq T$ and $R' \subseteq R$ satisfy the following conditions:
\begin{enumerate}
\item $T'$ contains no foreign key attributes involved in an entity reference from $T$ or multivalued reference to $T$.\footnote{If attributes excluded from $T'$ by this condition are in the primary key of $T$, then we will underline any corresponding entity references in $R'$.  (This does not apply to entity references excluded from $R'$ by condition 3.)}
\item A multivalued reference to some table $U$ can be included in $R'$ only if there exists no other abstract table $A(V) = (V', R_V')$ where $R_V'$ contains a multivalued reference to $U$.
\item For any $A(W) = (W', R_W')$ where $W$ is an entity table, an entity reference $r$ from $T$ to $W$ can be included in $R'$ only if $r$ is not a redundant entity reference corresponding to a multivalued reference in $R_W'$.
\end{enumerate}
\end{definition}
We will write $A(T) = (\{a_1, \ldots ,a_n\}, \{r_1, \ldots ,r_n\})$ in a shorthand:
\begin{eqnarray*}
A(T) & : & a_1, \ldots ,a_n, r_1, \ldots ,r_n
\end{eqnarray*}
where $a_1, \ldots ,a_n, r_1, \ldots ,r_n$ may be arranged in any order.

Our example can now be written as:
\begin{eqnarray*}
A(E) & : & \underline{emp}, \varepsilon(empdept, D), \mu(empphone, P) \\
A(D) & : & \underline{div, dept}, addr \\
A(P) & : & \underline{phone}
\end{eqnarray*}

If there exists a set of tables $X = \{T_1, \ldots ,T_n\}$ and corresponding $A(T_i) = (T_i', R_i')$ where $T_i \in X$, we may consider a directed graph with each vertex $v$ representing a table $T_i$ and arcs directed from $v$ representing entity references and multivalued references in $R_i'$.  This graph may have directed cycles.  The smallest possible directed cycle is formed by a single entity table containing a recursive entity reference, such as an employee having a manager that is also an employee, e.g.\ $A(E) : \underline{emp}, \varepsilon(empmgr, E)$.  As an example with two tables, if an employee can report to multiple managers, we might have $A(E) : \underline{emp}, \mu(empmgr, M)$ and $A(M) : \underline{\varepsilon(mgr, E)}$.

\section{Attributes}
In designing a new table, we will discuss attributes at a more abstract level than attributes in the relational model.  We ask two questions about each new abstract attribute:
\begin{enumerate}
\item \emph{Repeating}:  Is the attribute allowed to contain a \emph{Single} (exactly one) value or \emph{Multiple} (zero or more) values?  [$S$ or $M$]
\item \emph{Defining}:  Is the attribute domain \emph{Nonentity} (defined by the database system as a type or domain\footnote{For example, \texttt{CHARACTER} or \texttt{INTEGER} in SQL.}) or \emph{Entity} (defined by a table)?  [$N$ or $E$]
\end{enumerate}
The characteristics, \emph{repeating} and \emph{defining}, are independent of each other.  $S$ and $M$ are mutually exclusive, and $N$ and $E$ are mutually exclusive, in the sense that only one of each characteristic may be chosen.

The choice between $N$ and $E$ involves a few considerations.  An $E$ attribute has a domain defined by the elements of a table.  Each tuple of the defining table represents one element of the domain.  Conceptually, an $E$ attribute references an independent entity which can have its own properties via additional attributes.  The entity can also be referenced by other tables.  For example, in the previous section, employees are defined as working only in a department contained in the department table.  In that case, the employee department is an $E$ attribute, and the department address is a property of the department entity defined in the department table.  Other tables in addition to the employee table could reference the department table.  On the other hand, an $N$ attribute is not constrained by the values of another table, and there should be no attributes that describe its properties.  In the previous example, a department's address is probably an $N$ attribute because it is unconstrained (assuming there is no universal database of possible addresses) and does not have properties of its own defined by other attributes.  It probably also will not need to be used as a domain by other tables.

The choice of $M$ includes cases where an attribute is \emph{optional} (allowing either one value or no value).  Selecting $M$ for an optional attribute is preferred over using an $S$ attribute with null values.  Also, $M$ is intended mainly for cases in which a variable number of values is possible.  If the number of values is constant, then it may be preferable to create a fixed number of $S$ attributes.

When the abstract attributes are translated to the relational model, these characteristics determine the number of additional tables that will be needed (see \emph{Table~1}) and the placement of foreign keys.

\begin{table}
\centering
\caption{Number of database tables required}
\begin{tabular}{ | l | c | c | }
\hline
 & \textbf{Nonentity (N)} & \textbf{Entity (E)} \\
\hline
\textbf{Single (S)} & 0 & 1 \\
\hline
\textbf{Multiple (M)} & 1 & 2 \\
\hline
\end{tabular}
\end{table}

\section{Data model}
This section starts with a new example based on data collection in evolutionary biology research.  Suppose that one wants to design a database to record individual specimens of various organisms collected in the field, and for each individual someone has taken photographic images and recorded its organism type (such as genus and species).  One way to begin would be to define an entity table representing the individuals.  In order to allow for hybrids, we will store ``biotypes'' defined as combinations of organism types.  Both biotypes and organism types will be modeled as entities, to allow for the possibility that one may want to describe them with additional attributes.  We can express the ``individual'' entity $I$ as, ``An individual in $I$ is uniquely identified by a name $indname$ (a Single Nonentity attribute), and has photo images $images$ (a Multiple Nonentity attribute) and a biotype $biotype$ (a Single Entity attribute).''  $MN$ attributes are notated using the $\mu()$ operator, and $SE$ attributes are notated with $\varepsilon()$:
\begin{eqnarray*}
A(I) & : & \underline{indname}, \mu(images, G), \varepsilon(biotype, B)
\end{eqnarray*}

This implies two new tables, $G$ and $B$.  We proceed to defining the multivalued table $G$:  ``For each individual in $I$, $G$ contains a set of unique image file names $imgfile$ (a Single Nonentity attribute), each having an associated comment $notes$ (a Single Nonentity attribute).''  This is notated as:
\begin{eqnarray*}
A(G) & : & \underline{imgfile}, notes
\end{eqnarray*}

The entity table $B$ is defined as:  ``A biotype in $B$ is uniquely identified by a name $btname$ (a Single Nonentity attribute), and has organisms $orgs$ (a Multiple Entity attribute).''  $ME$ attributes are notated by combining $\mu()$ and $\varepsilon()$:
\begin{eqnarray*}
A(B) & : & \underline{btname}, \mu(orgs, T) \\
A(T) & : & \underline{\varepsilon(org, O)}
\end{eqnarray*}

Next, the entity table $O$ must be defined:  ``An organism in $O$ is uniquely identified by a genus $genus$ (a Single Nonentity attribute) and a species $species$ (a Single Nonentity attribute), and has a common name $cname$ (a Single Nonentity attribute).''  Therefore, including the entire example:
\begin{eqnarray*}
A(I) & : & \underline{indname}, \mu(images, G), \varepsilon(biotype, B) \\
A(G) & : & \underline{imgfile}, notes \\
A(B) & : & \underline{btname}, \mu(orgs, T) \\
A(T) & : & \underline{\varepsilon(org, O)} \\
A(O) & : & \underline{genus, species}, cname
\end{eqnarray*}

This would be implemented in a database as:
\begin{eqnarray*}
I & = & \{ \underline{indname}, btname \} \\
G & = & \{ \underline{indname, imgfile}, notes \} \\
B & = & \{ \underline{btname} \} \\
T & = & \{ \underline{btname, genus, species} \} \\
O & = & \{ \underline{genus, species}, cname \} \\
images & = & \Phi_{G,I}(\{indname\}) \\
biotype & = & \Phi_{I,B}(\{btname\}) \\
orgs & = & \Phi_{T,B}(\{btname\}) \\
org & = & \Phi_{T,O}(\{genus, species\})
\end{eqnarray*}

The same relational schema structure would result if organisms were defined as participating in multiple biotypes:
\begin{eqnarray*}
A(O) & : & \underline{genus, species}, cname, \mu(biotypes, T) \\
A(T) & : & \underline{\varepsilon(biotype, B)} \\
A(B) & : & \underline{btname} \\
\mathrm{etc.}
\end{eqnarray*}

Note that this schema is in 4NF, and redundancies have been avoided by removing multivalued attributes to new tables using $\mu()$.  Returning to the first example in this paper, it is natural to see that an employee's phone numbers and list of publications are unrelated, which might suggest $A(E) : \underline{emp}, \mu(empphone,$ $P), \mu(emppub, B)$ where $B$ is a table representing an employee's publications.  The less desirable $A(E) : \underline{emp}, phones, pubs$ (probably not in 1NF) and $A(E) : \underline{emp, phone, pub}$ (not in 4NF) are possible but not as intuitive.

The use of entity references may help to clarify the role of attributes.  The problems addressed by 2NF and 3NF can be avoided if attributes that need to be described by other attributes are removed to new entity tables using $\varepsilon()$.

This paper suggests that abstracting entity references and multivalued references, while otherwise remaining close to the relational model, can promote intuitive normalization.

\section{Hierarchical application}
If there exists a set of tables $X = \{T_1, \ldots ,T_n\}$ and corresponding $A(T_i) = (T_i', R_i')$ where $T_i \in X$, we may consider a directed graph with each vertex $v$ representing a table $T_i$ and arcs directed from $v$ representing multivalued references in $R_i'$.  This graph is a set of trees or hierarchies with entity tables as the root nodes.  We rewrite an abstract table $A(T)$ in this hierarchical form using a shorthand denoted by $H(T)$, and we include any multivalued table as contained within its parent.  If we indicate multivalued tables with parentheses and entity references with square brackets, then the biology example from the previous section might be described as:
\begin{eqnarray*}
H(I) & : & \underline{indname}, images (\underline{imgfile}, notes), biotype [B] \\
H(B) & : & \underline{btname}, orgs (\underline{org [O]}) \\
H(O) & : & \underline{genus, species}, cname
\end{eqnarray*}
A larger example shown in \emph{Figure~1} describes part of a database for high throughput genetic sequencing research.\footnote{The database was designed by Chris Bizon et al.}  Here multivalued tables are drawn as nested tables and entity references are italicized.  (Primary keys are in boldface.)  \emph{Figure~2} depicts the same schema in a conventional format.

\begin{figure}[h]
\centering
\includegraphics[width=4.75in]{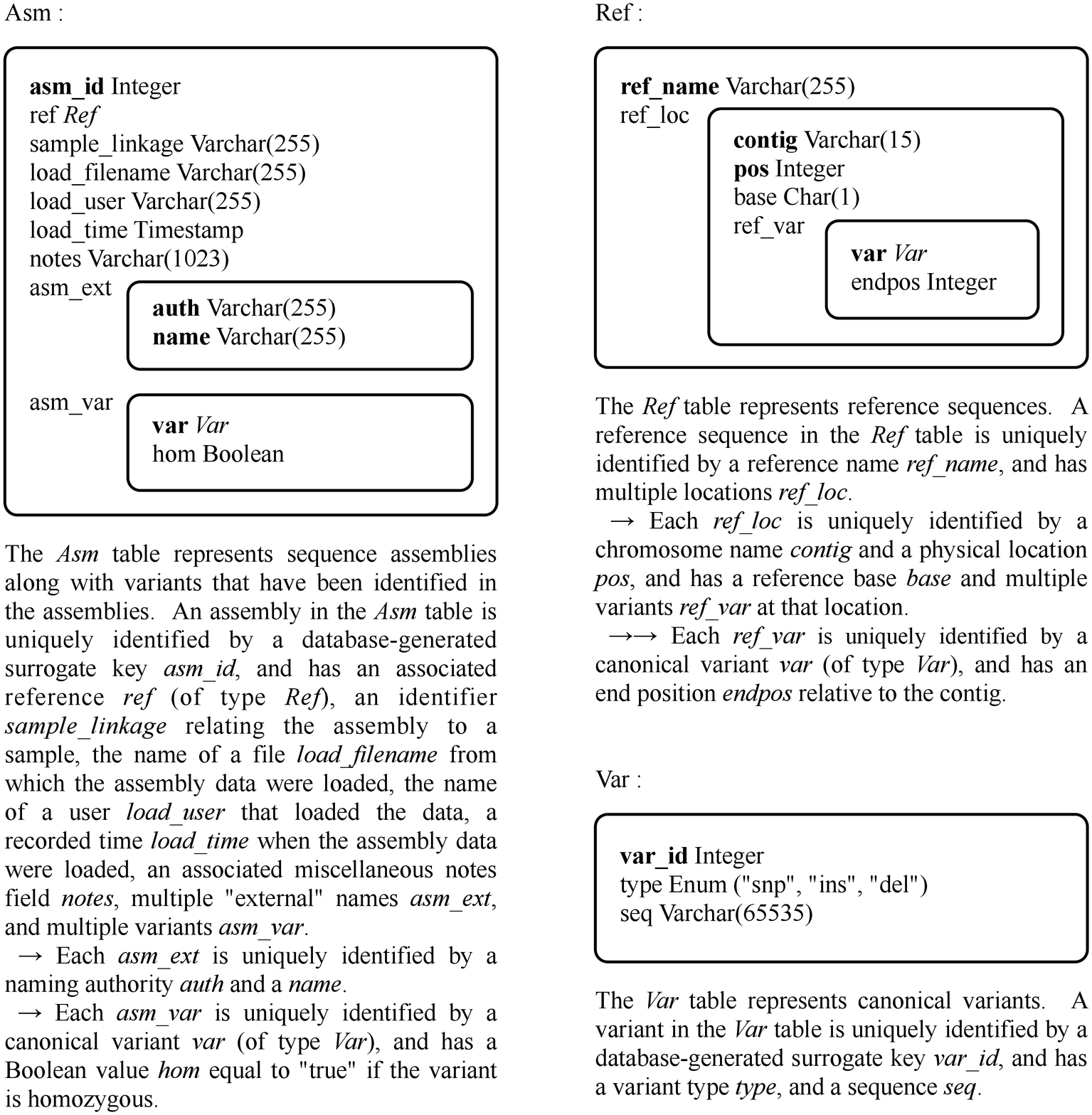}
\caption{Sequence database, with abstract attributes and hierarchical tables}
\end{figure}

\begin{figure}[h]
\centering
\includegraphics[width=4.75in]{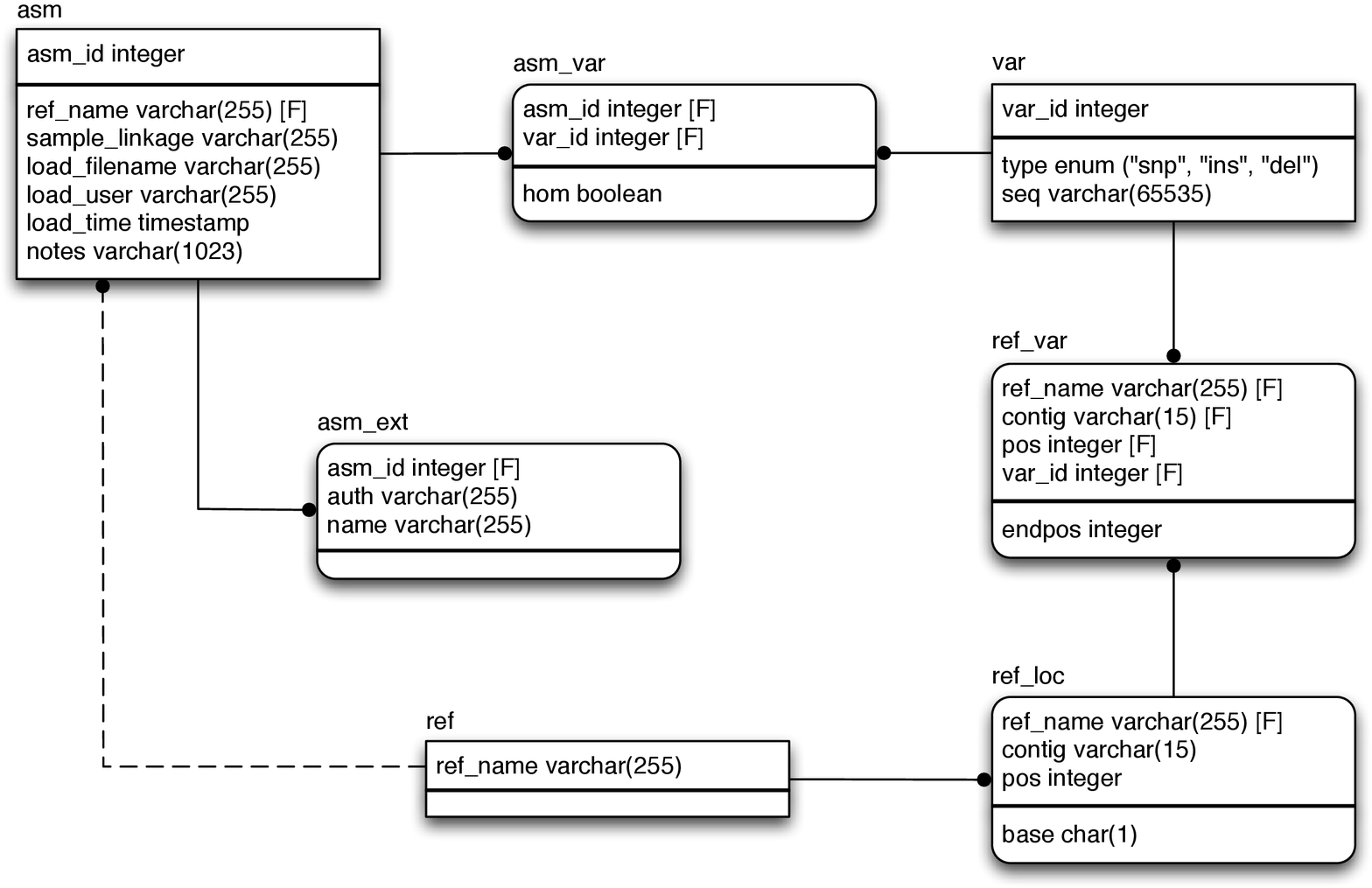}
\caption{Sequence database, entity-relationship logical model (IDEF1X)}
\end{figure}

\section{Conclusions}
The paper described an alternative method of data modeling intended for non-specialists, based on a simple transformation of relational schemas.

\bibliographystyle{abbrv}
\bibliography{nassar_draft}

\begin{thebibliography}{1}

\bibitem{bell09}
G.~Bell, T.~Hey, and A.~Szalay.
\newblock Beyond the data deluge.
\newblock {\em Science}, 323(5919):1297--1298, 2009.

\bibitem{chen76}
P.~P. Chen.
\newblock The entity-relationship model---toward a unified view of data.
\newblock {\em ACM Trans. Database Syst.}, 1(1):9--36, 1976.

\bibitem{codd71}
E.~F. Codd.
\newblock Further normalization of the data base relational model.
\newblock {\em IBM Research Report, San Jose, California}, RJ909, 1971.

\bibitem{davies06}
I.~Davies, P.~Green, M.~Rosemann, M.~Indulska, and S.~Gallo.
\newblock How do practitioners use conceptual modeling in practice?
\newblock {\em Data Knowl. Eng.}, 58(3):358--380, 2006.

\bibitem{tavana07}
M.~Tavana, P.~Joglekar, and M.~A. Redmond.
\newblock An automated entity-relationship clustering algorithm for conceptual
  database design.
\newblock {\em Inf. Syst.}, 32(5):773--792, 2007.

\end{thebibliography}

\end{document}